%% file: Specs-Base.tex
\documentclass{llncs}

\pdfoutput=1

\renewcommand{\baselinestretch}{0.99}

 \usepackage{scalerel}
\input{Auxiliaries/Package-Imports.tex}

\input{Auxiliaries/Commands.tex}

\input{Auxiliaries/Specs.tex}

\title{Symbolic Abstract Contract Synthesis in a Rewriting Framework \thanks{This work has been partially supported by the EU (FEDER)
and Spanish MINECO under grants
TIN2015-69175-C4-1-R
and TIN2013-45732-C4-1-P, and
by Generalitat Valenciana  PROMETEOII/2015/013. Daniel Pardo was supported by FPU-ME grant FPU14/01830.}
}
\author{Mar\'{\i}a Alpuente \inst{1} \and Daniel Pardo\inst{1} \and
  Alicia Villanueva\inst{1}}
\institute{DSIC, Universitat Polit\`ecnica de Val\`encia\\
Camino de Vera s/n, 46022 Valencia, Spain\\
\email{\{alpuente,daparpon,villanue\}@dsic.upv.es}}

\renewcommand{\implies}{\Rightarrow}

\begin{document}
\pagestyle{headings} 
\maketitle

\begin{abstract}
    We propose an automated technique for inferring
    software contracts from programs that are written in a non-trivial
    fragment of \C, called \KernelC, that  supports
      pointer-based structures and heap manipulation. Starting from the semantic
    definition of \KernelC\ in the \K\ framework, we enrich the
    symbolic execution facilities recently provided by \K\ with
    novel capabilities for assertion synthesis that are based on abstract subsumption. Roughly
    speaking, we define an abstract symbolic technique that explains
    the execution of a (modifier) \C\ function  by using other (observer) routines in the
    same program. We implemented our technique in the automated tool
    \KindSpec, which generates logical axioms that express pre- and
    post-condition assertions by defining the precise input/output
    behavior of the  \C\ routines.
\end{abstract}

  \vspace{-0.5ex}
\section{Introduction}\label{sec:intro}
\vspace{-0.5ex}
\input{Sections/Introduction.tex}

\vspace{-2.75ex}
\section{Method Specification: A Running Example}\label{sec:example}
\vspace{-.75ex}
\input{Sections/Running-Example.tex}

\vspace{-2ex}
\section{The (symbolic) \K  Framework}\label{sec:k}
\vspace{-1ex}
\input{Sections/KFramework.tex}

\vspace{-2ex}
\section{Improving Symbolic Execution in \K\ }\label{sec:symbexec}
\vspace{-1ex}
\input{Sections/Symbolic-Execution.tex}

\vspace{-1ex}
\section{Inference Algorithm}\label{sec:inference}
\vspace{-.5ex}
\input{Sections/Inference.tex}
\vspace{-2ex}
\section{Related work and Conclusions}\label{sec:conc-rel}
\vspace{-1ex}
\input{Sections/RelatedConc.tex}
\vspace{-1ex}

\renewcommand{\baselinestretch}{0.98}

\input{small-bbl2.bbl}

\appendix
\clearpage
\input{Sections/Appendix.tex}

\end{document}

%% file: Auxiliaries/Package-Imports.tex
\usepackage[american]{babel}
\usepackage{mathtools}
\usepackage{marginnote}
\usepackage{color}
\usepackage{graphicx}
\usepackage{algorithm,algorithmic}
\usepackage{listings}			
\usepackage{paralist}			
\usepackage{multicol}
\usepackage[style=math]{Styles/k}
\usepackage{tikz}
\usetikzlibrary{backgrounds, shapes, fit}
\usepackage{smartref}

%% file: Auxiliaries/Commands.tex
\makeatletter
\def\ifempty#1{%
    \if\relax\detokenize{#1}\relax%
        \expandafter\@firstoftwo
    \else
        \expandafter\@secondoftwo
    \fi}
\makeatother

\newcommand{\cellface}[1]{{\color{white!50!black}\sf #1}}
\newcommand{\heap}{\cellface{heap}}

\newcommand{\env}{\cellface{env}}
\newcommand{\kcompcell}{\cellface{k}}
\newcommand{\pathcondcell}{\cellface{path-condition}}

\newcommand{\initheap}{\cellface{init-heap}}
\newcommand{\stack}{\cellface{stack}}

\newcommand{\tv}{\operatorname{tv}}

\newcommand{\KindSpec}{\mbox{\sc KindSpec 2.0}}
\newcommand{\C}{\mbox{\sc C}}
\newcommand{\KernelC}{\mbox{\sc KernelC}}
\newcommand{\MatchC}{\mbox{\sc MatchC}}
\newcommand{\QuickSpec}{\mbox{\sc QuickSpec}}

\newcommand{\SE}{\textsc{SE}}
\newcommand{\SEAS}{$\textsc{SE}^\sharp$}

\newcommand{\symba}[1]{\text{\sf\&{#1}}}

\newcommand{\symbv}[1]{\text{\sf?{#1}}}

\newcommand{\mapsTo}[2]{#1\mapsto{}#2}
\newcommand{\ie}{i.e.,}

\newcommand{\explain}[1]{\ifempty{#1}{\mathop{explain}}{\mathop{explain}(#1)}}


%% file: Auxiliaries/Specs.tex
\newcommand{\SpecExpectedRefined}{
\begin{align*}
\left(
\begin{array}{l}
\mathtt{isnull(s)}=\mathtt{1}
\end{array}
\right)
& \Rightarrow
\left(
\begin{array}{l}
\mathtt{isnull(s')}=\mathtt{1} \; \wedge \; 
\mathtt{ret}=\mathtt{0}
\end{array}
\right)
\\
\left(
\begin{array}{l}
\mathtt{isfull(s)}=\mathtt{1}
\end{array}
\right)
& \Rightarrow
\left(
\begin{array}{l}
\mathtt{isfull(s')}=\mathtt{1} \; \wedge \; 
\mathtt{contains(s',x)}=\mathtt{contains(s,x)} \; \wedge \\
\mathtt{length(s')}=\mathtt{length(s)} \; \wedge\; 
\mathtt{ret}=\mathtt{0}
\end{array}
\right)
\\
\left(
\begin{array}{l}
\mathtt{contains(s,x)}=\mathtt{1}
\end{array}
\right)
& \Rightarrow
\left(
\begin{array}{l}
\mathtt{contains(s',x)}=\mathtt{1} \; \wedge \; 
\mathtt{length(s')}=\mathtt{length(s)} \; \wedge \; 
\mathtt{ret}=\mathtt{0}
\end{array}
\right)
\\
\left(
\begin{array}{l}
\mathtt{isempty(s)}=\mathtt{1} \; \wedge 
\mathtt{isfull(s)}=\mathtt{0}
\end{array}
\right)
& \Rightarrow
\left(
\begin{array}{l}
\mathtt{isempty(s')}=\mathtt{0} \; \wedge \; 
\mathtt{contains(s',x)}=\mathtt{1} \; \wedge \\
\mathtt{length(s')}=\mathtt{length(s)}+1 \; \wedge \; 
\mathtt{ret}=\mathtt{1}
\end{array}
\right)
\\
\left(
\begin{array}{l}
\mathtt{isnull(s)}=\mathtt{0} \; \wedge 
\mathtt{isempty(s)}=\mathtt{0} \; \wedge \\
\mathtt{isfull(s)}=\mathtt{0} \; \wedge 
\mathtt{contains(s,x)}=\mathtt{0}
\end{array}
\right)
& \Rightarrow
\left(
\begin{array}{l}
\mathtt{isnull(s')}=\mathtt{0} \; \wedge \; 
\mathtt{isempty(s')}=\mathtt{0} \; \wedge \; \\
\mathtt{contains(s',x)}=\mathtt{1} \: \wedge \; 
\mathtt{length(s')}=\mathtt{length(s)}+1 \; \wedge \\
\mathtt{ret}=\mathtt{1}
\end{array}
\right)
\end{align*}
}

\newcommand{\SpecActualSubsumption}{
\begin{align*}
{\tt A1} & \left(
\begin{array}{l}
\mathtt{isnull(s)}=\mathtt{1} \; \wedge 
\mathtt{isempty(s)}=\mathtt{0} \; \wedge \\
\mathtt{isfull(s)}=\mathtt{0} \; \wedge 
\mathtt{contains(s,x)}=\mathtt{0} \; \wedge \\
\mathtt{length(s)}=\mathtt{0}
\end{array}
\right)
& \Rightarrow &
\left(
\begin{array}{l}
\mathtt{isnull(s')}=\mathtt{1} \; \wedge 
\mathtt{isempty(s')}=\mathtt{0} \; \wedge \\
\mathtt{isfull(s')}=\mathtt{0} \; \wedge 
\mathtt{contains(s',x)}=\mathtt{0} \; \wedge \\
\mathtt{length(s')}=\mathtt{0} \; \wedge 
\mathtt{ret}=\mathtt{0}
\end{array}
\right)
\\
{\tt A2} & \left(
\begin{array}{l}
\mathtt{isnull(s)}=\mathtt{0} \; \wedge 
\mathtt{isempty(s)}=\_i1 \; \wedge \\
\mathtt{isfull(s)}=\mathtt{1} \; \wedge 
\mathtt{contains(s,x)}=\_i2 \; \wedge \\
\mathtt{length(s)}=\_i3
\end{array}
\right)
& \Rightarrow &
\left(
\begin{array}{l}
\mathtt{isnull(s')}=\mathtt{0} \; \wedge 
\mathtt{isempty(s')}=\_i1 \; \wedge \\
\mathtt{isfull(s')}=\mathtt{1} \; \wedge 
\mathtt{contains(s',x)}=\_i2 \; \wedge \\
\mathtt{length(s')}=\_i3 \; \wedge 
\mathtt{ret}=\mathtt{0}
\end{array}
\right)
\\
{\tt A3} & \left(
\begin{array}{l}
\mathtt{isnull(s)}=\mathtt{0} \; \wedge 
\mathtt{isempty(s)}=\mathtt{0} \; \wedge \\
\mathtt{isfull(s)}=\mathtt{0} \; \wedge 
\mathtt{contains(s,x)}=\mathtt{1} \; \wedge \\
\mathtt{length(s)}=\_i1
\end{array}
\right)
& \Rightarrow &
\left(
\begin{array}{l}
\mathtt{isnull(s')}=\mathtt{0} \; \wedge 
\mathtt{isempty(s')}=\mathtt{0} \; \wedge \\
\mathtt{isfull(s')}=\mathtt{0} \; \wedge 
\mathtt{contains(s',x)}=\mathtt{1} \; \wedge \\
\mathtt{length(s')}=\_i1 \; \wedge 
\mathtt{ret}=\mathtt{0}
\end{array}
\right)
\\
{\tt A4 } & \left(
\begin{array}{l}
\mathtt{isnull(s)}=\mathtt{0} \; \wedge 
\mathtt{isempty(s)}=\mathtt{1} \; \wedge \\
\mathtt{isfull(s)}=\mathtt{0} \; \wedge 
\mathtt{contains(s,x)}=\mathtt{0} \; \wedge \\
\mathtt{length(s)}=\mathtt{0}
\end{array}
\right)
& \Rightarrow &
\left(
\begin{array}{l}
\mathtt{isnull(s')}=\mathtt{0} \; \wedge 
\mathtt{isempty(s')}=\mathtt{0} \; \wedge \\
\mathtt{contains(s',x)}=\mathtt{1} \; \wedge 
\mathtt{length(s')}=\mathtt{1} \; \wedge \\
\mathtt{ret}=\mathtt{1}
\end{array}
\right)
\\
{\tt A5} & \left(
\begin{array}{l}
\mathtt{isnull(s)}=\mathtt{0} \; \wedge 
\mathtt{isempty(s)}=\mathtt{0} \; \wedge \\
\mathtt{isfull(s)}=\mathtt{0} \; \wedge 
\mathtt{contains(s,x)}=\mathtt{0} \; \wedge \\
\mathtt{length(s)}=\mathtt{1}
\end{array}
\right)
& \Rightarrow &
\left(
\begin{array}{l}
\mathtt{isnull(s')}=\mathtt{0} \; \wedge 
\mathtt{isempty(s')}=\mathtt{0} \; \wedge \\
\mathtt{contains(s',x)}=\mathtt{1} \; \wedge 
\mathtt{length(s')}=\mathtt{2} \; \wedge \\
\mathtt{ret}=\mathtt{1}
\end{array}
\right)
\\
{\tt A6} & \left(
\begin{array}{l}
\mathtt{isnull(s)}=\mathtt{0} \; \wedge 
\mathtt{isempty(s)}=\mathtt{0} \; \wedge \\
\mathtt{isfull(s)}=\mathtt{0} \; \wedge 
\mathtt{contains(s,x)}=\mathtt{0} \; \wedge \\
\mathtt{length(s)}=\mathtt{2}
\end{array}
\right)
& \Rightarrow &
\left(
\begin{array}{l}
\mathtt{isnull(s')}=\mathtt{0} \; \wedge 
\mathtt{isempty(s')}=\mathtt{0} \; \wedge \\
\mathtt{contains(s',x)}=\mathtt{1} \; \wedge 
\mathtt{length(s')}=\mathtt{3} \; \wedge \\
\mathtt{ret}=\mathtt{1}
\end{array}
\right)
\\
{\tt A7} & \left(
\begin{array}{l}
\mathtt{isnull(s)}=\mathtt{0} \; \wedge 
\mathtt{isempty(s)}=\mathtt{0} \; \wedge \\
\mathtt{isfull(s)}=\mathtt{0} \; \wedge 
\mathtt{contains(s,x)}=\mathtt{0} \; \wedge \\
\mathtt{length(s)}=\mathtt{3}
\end{array}
\right)
& \Rightarrow &
\left(
\begin{array}{l}
\mathtt{isnull(s')}=\mathtt{0} \; \wedge 
\mathtt{isempty(s')}=\mathtt{0} \; \wedge \\
\mathtt{contains(s',x)}=\mathtt{1} \; \wedge 
\mathtt{length(s')}=\mathtt{4} \; \wedge \\
\mathtt{ret}=\mathtt{1}
\end{array}
\right)
\\
{\tt C1} & \left(
\begin{array}{l}
\mathtt{isnull(s)}=\mathtt{0} \; \wedge 
\mathtt{isempty(s)}=\mathtt{0} \; \wedge \\
\mathtt{isfull(s)}=\mathtt{0} \; \wedge 
\mathtt{contains(s,x)}=\_i1 \; \wedge \\
\mathtt{length(s)}=\_i2
\end{array}
\right)
&\Rightarrow &
\left(
\begin{array}{l}
\mathtt{isnull(s')}=\mathtt{0} \; \wedge 
\mathtt{isempty(s')}=\mathtt{0} \; \wedge \\
\mathtt{contains(s',x)}=\mathtt{1} \; \wedge 
\mathtt{length(s')}=\_i2+1 \; \wedge \\
\mathtt{ret}=\mathtt{1}
\end{array}
\right)
\end{align*}
}


%% file: Sections/Introduction.tex

Checking software contracts 
\cite{meyer_applying_1992} is 
one
  of the most promising techniques for achieving software
  reliability. Contracts essentially consist of requirements that are imposed on the
  arguments and   result values    when   functions are invoked.  Given its
  interest, considerable 
effort has recently been invested towards
  giving automatic support for equipping programs with extensive contracts,
  yet the current contract inference tools are still often
  unsatisfactory in practice \cite{Cousot2013}. 

  This paper describes a symbolic inference system  that
  synthesizes 
  contracts for heap-manipulating programs that are written in a
  non-trivial fragment of \C,\ called \KernelC\
  \cite{ellison_executable_2012}, 
which includes functions, I/O primitives, dynamically
  allocated structures,  and pointer manipulation.  By automating the tedious and time-consuming process of generating
  contracts, programmers can reap the benefits of assertion--based
  debugging and verification methods with reasonable effort.

Given a program $P$, the 
contract discovery problem
is generally 
 described as the problem of inferring a likely
 specification for every function $m$ in $P$ that uses I/O primitives
 and/or modifies the state.
The specifications that we aim to infer consist of logical 
  assertions   that characterize
  the function behavior 
  and that are expressed as method
  pre-conditions (imposed on the arguments) and post-conditions (relating the
  arguments and the result for a method). 
 
In
\cite{AFV2013},  a preliminary specification inference  technique 
 was proposed that is based on  
 the classification scheme for data abstractions
developed in \cite{liskov_abstraction_1986}, 
where a
function (method) may be either a {\it constructor}, a {\it modifier}, or an 
{\it observer}. The intended behavioral specification
 of any \emph{modifier} function $m$  of $P$ is expressed as
a set of logical assertions that characterize   the pre- and post-states  of the $m$ execution by using the
 \emph{observer} functions in $P$.
For instance, for  the case of 
 a modifier function {\tt push} that
 adds the element {\tt x} into a given bounded stack {\tt q} (and assuming the traditional meaning for the observer functions {\tt top}, {\tt isfull}, and {\tt size}), a
 typically expected axiom could be {\tt   isfull(q)=0  $\wedge$ size(q)=n $\implies$ top(q)=x $\wedge$ size(q)=n+1}.
 
The inference technique of \cite{AFV2013} relies  on symbolic execution (SE) \cite{king_symbolic_1976},
a well-known program analysis technique that allows  programs to be
executed using \emph{symbolic} input values instead of actual
(concrete) data so that the  program   execution manipulates symbolic
expressions involving the symbolic values. 
 More precisely, for each pair ($s$,$s'$) of initial and
final states 
in the symbolic execution of {\em m},
an implicative axiom $p \implies q$ is synthesized where both the 
antecedent  $p$ and the consequent $q$  are
expressed in terms of the (sub-)set of program observers that \emph{explain} 
$s$ and $s'$. 
This is achieved by analyzing the results of   symbolically executing each observer method $o$
  from  initial configurations that contain a 
  symbolic characterization of
$s$ and $s'$.
The symbolic  infrastructure 
employed in  \cite{AFV2013} 
 was  built on top of  the
  (rewriting logic) semantic framework \K,\  which greatly  facilitates the development of
 executable semantics  of programming languages and related formal 
analysis techniques \cite{rosu_overview_2010}. 
Unfortunately,  
it was developed by reusing spare features  
 of a formal verifier for \KernelC\ (called \MatchC) 
 that was  
formerly provided   within \K\  
but is currently unsupported. On the other hand, 
the underlying methodology  in  \cite{AFV2013} was rather limited 
since a fixed threshold for   loop unrolling was  imposed  in order to avoid  
 non-termination risks.
In \cite{AFV2015},  we switched to a recent, native   extension of \K\ 
that  supports symbolic execution by language transformation  \cite{arusoaie_symbolic_2015}. 
 However, the methodology in \cite{AFV2015} inherited
the   loop unrolling strategy based on depth bounds from \cite{AFV2013}.

In this work, we improve the inference power of  \cite{AFV2013,AFV2015} by endowing \K's\ symbolic execution with modern subsumption techniques based on approximation \cite{anand_symbolic_2008} and 
lazy initialization \cite{khurshid_generalized_2003}. The fact that this   symbolic infrastructure is  
much more flexible and (potentially) language-independent allows us to define a generic,
more accurate, easily maintainable and robust framework for the inference of program contracts that could be   adapted to other languages
defined within the \K\ framework with negligible effort. 
We summarize our contributions as follows.

\vspace{-0ex}
\begin{enumerate}
\item A symbolic algorithm that synthesizes 
contracts for heap-manipulating code while coping with infinite
computations. This is done by 

\begin{enumerate}
\item augmenting
  \K's\ symbolic execution with lazy initialization and 
 a widening operator   based on abstract subsumption (in Section \ref{sec:symbexec}),
 and 
 \item synthesizing method pre- and post-conditions by means of a contract  inference algorithm that explains the (initial and final) abstract symbolic execution states by using the program observers (in Section \ref{sec:inference}).
 \end{enumerate}

Because of the abstraction, 
some inferred axioms for method $m$ cannot be guaranteed to be   correct 
and are kept apart as ``candidate'' (or overly general) axioms.
A contract refinement algorithm is then formalized that tries to falsify  them
 by  checking whether an input call to $m$
 that satisfies the 
 axiom antecedent ends in a final state that does not satisfy the given consequent.

\item  The proposed inference technique is
  implemented  in the \textsc{KindSpec} 2.0 system, which 
  builds on the capabilities of the SMT solver Z3 \cite{moura_z3:_2008} to simplify the axioms. 
Also, the inferred contracts are given a compact representation
that abstracts the user from any implementation details.

\end{enumerate}

%% file: Sections/Running-Example.tex
By abuse, we use the standard  terminology for contracts of
object-oriented programming  and speak of {\em methods} when we refer
to \KernelC\    functions.
 Like many state-of-the-art formal specification approaches, we assume
 to be working in
 a contract-based setting \cite{meyer_applying_1992}, where the granularity of specification units is at
 the level of one method. 
  Our  inference technique  
 relies on 
  the classification scheme
developed for data abstractions in \cite{liskov_abstraction_1986}, 
where a
function (method) may be either a {\it constructor}, a {\it modifier}, or an 
{\it observer}. A constructor returns a new data object 
 from scratch (i.e., without taking the object as an input parameter).  A
modifier alters an existing 
object (i.e., it changes the state
of one or more of the data attributes in the instance).  An observer
inspects the object and returns a value characterizing one or more of
its state attributes. 
Since the \C\ language does not enforce data encapsulation, we cannot
assume purity of any function; thus, we do not assume the traditional
premise 
that states that observer
functions  do not cause side effects. 
 In other words, any function can potentially be a
 modifier 
   and we simply define an observer as any function whose
 return type is different from \texttt{void}.

Let us introduce the leading example that we use to describe our
inference methodology: 
a \KernelC\
implementation of an abstract datatype for representing sets by
using linked lists. The example is composed of 7 methods:
one constructor (\texttt{new}), one modifier (\texttt{insert}), and
five observers (\texttt{isnull}, \texttt{isempty}, \texttt{isfull},
\texttt{contains}, and \texttt{length}). 
 Note that the observers  in 
this program  do not modify any program
objects, even if 
purity of observers is not required in our framework.
As  is usual in \C,\ logical observers return 
value {\tt 1} (resp.\ {\tt 0}) to represent the traditional boolean value 
\emph{true} (resp.\ \emph{false}).

\begin{example}\label{ex:program}
Consider the program fragment 
given in Figure~\ref{fig:code} (the full program code can be found in Appendix~\ref{apndx:full_program}), where we define set operations over a data structure 
  ({\tt struct set}) that records the number of elements  contained in the set (field {\tt size}), the maximum number of elements that  can be held (field {\tt capacity}), and a pointer to a list that stores the   set elements (field {\tt elems}).
  Each node of the list is a record  data structure ({\tt struct lnode}) that contains an integer value (field {\tt value}) and a pointer to the subsequent list element  (field {\tt next}).

\begin{figure}[ht!]
\lstset{
language=C,
basicstyle={\ttfamily\scriptsize},
breaklines=true,
numbers=left,
stepnumber=1,
tabsize=1,
columns=[l|l]flexible
}
\setlength\columnsep{15pt}
\begin{multicols}{2}
\begin{lstlisting}
#include <stdlib.h>

struct lnode{
	int value;
	struct lnode *next;
};
struct set {
	int capacity;
	int size;
	struct lnode *elems;
};
struct set* new(int capacity) {...}

int insert(struct set *s, int x) {
	struct lnode *new_node;
	struct lnode *end_node;
	struct lnode *n;
	
	if(s==NULL)
		return 0; /* NULL set */		
	if(s->size >= s->capacity)
		return 0; /* no space left */
	
	end_node = s->elems;	
	n = end_node;
	while(n != NULL) {
		if(n->value == x)
			return 0; /* x already in the set */
		end_node = n;
		n = n->next;
	}
	
	new_node = (struct lnode*) malloc(sizeof(struct lnode));
	if(new_node == NULL)
		return 0; /* no memory left */
	new_node->value = x;
	new_node->next = s->elems;
	
	s->elems = new_node;
	s->size += 1;
	return 1; /* element added */
}

int isnull(struct set *s) {...}
int isempty(struct set *s) {...}
int isfull(struct set *s) {...}
int contains(struct set *s, int x) {...}
int length(struct set *s) {...}
\end{lstlisting}
\end{multicols}
\caption{Fragment of the \KernelC\ implementation of a set datatype.}
\vspace{-3ex}
\label{fig:code}
\end{figure}

A call {\tt insert(s,x)} to the {\tt insert} function proceeds as
follows: it first checks that the pointer {\tt s} to the set structure is
different from {\tt NULL}, 
that the set is not full, 
and that {\tt x} is not in the set yet. 
Then, a new list node {\tt *new\_node} is allocated, filled with
the value {\tt x}, and inserted as the first element of the list.
 Also, 
the size of the set is increased by 1 and the call
returns {\tt 1}; 
otherwise, 
  {\tt 0} is returned and {\tt s} is not modified.

The following observers  return {\tt 0} unless explicitly stated otherwise. {\tt isnull(s)}
returns {\tt 1} if the pointer {\tt s} references to {\tt NULL}
memory; 
 {\tt isempty} returns {\tt 1} if {\tt s} is initialized but {\tt elems} is
{\tt NULL}; 
{\tt isfull(s)} returns {\tt 1} if the size
of {\tt s} is greater than or equal to its capacity; 
 and {\tt contains(s,x)} 
returns {\tt 1} if the
value {\tt x} is found in {\tt s}. 
The function {\tt length(s)} incrementally counts the number of
elements (nodes) in the set {\tt s} by traversing the list {\tt
  s->elems} and returns this number, or it returns {\tt 0} if the set {\tt s} pointer is {\tt NULL}.
\end{example}

From the source code of the program, for each
modifier function $m$, we aim to synthesize, a contract of the form $<P,Q,{\cal L}>$ where
$P$ is the method precondition, $Q$ is the method postcondition, and
${\cal L}$ is the set of program locations (local variables, data-structure pointers and fields, and method
parameters) that are (potentially) affected by the method
execution. We first compute
  a set of implication formulas $p \Rightarrow q$, where $p$ and $q$
  are conjunctions of equations $l = r$. The left-hand
  side $l$ of each equation can be either 1) a call to an observer
  function, and then $r$ represents the return value of that call; or
  2) the keyword \texttt{ret}, and then $r$ represents the value
  returned by the modifier function $m$ being observed.  
  Then, given the set of implication formulas $\{p_1 \Rightarrow q_1,
  \ldots, p_n \Rightarrow
  q_n\}$, 
    $P$ is defined as $p_1 \vee
    \ldots \vee p_n$, the postcondition $Q$ is the
    formula\footnote{This is similar to the idea of contracts with
      \emph{named behaviors} as provided in the ACSL contract
      specification language for \C\ \cite{baudin09acsl}.} $(p_1
    \Rightarrow q_1) \wedge \ldots \wedge (p_n \Rightarrow q_n)$, and
   the elements of  ${\cal L}$ refer
     to locations whose value
    might be affected by the execution of $m$, that is, 
    all memory
    locations of the pre-state that do not belong to the set ${\cal
      L}$ remain allocated and are left unchanged in the
    post-state. 
The set ${\cal L}$ itself is interpreted in the
    pre-state and 
is necessary for sound usage of contracts.

\begin{example}\label{ex:specification}
 The intended 
postcondition $Q$ 
for the modifier function {\tt insert(s,x)}
of   Example~\ref{ex:program} contains   five  
axioms (each one
  given by  an implication),
which are shown in Figure \ref{fig:expectedspec}.
We adopt the standard primed notation to distinguish variable values after the execution of  the method
from their value before the execution.

\begin{figure}[ht!]
\scriptsize
\SpecExpectedRefined
\vspace{-4ex}
{\caption{Expected postcondition axioms for the \texttt{insert} method}}
\label{fig:expectedspec}
\vspace{-4ex}

\end{figure}

The first axiom can be read as: if the outcome of
{\tt isnull(s)} is 1 before the call to 
{\tt insert(s,x)}, 
then, after execution, 
the set is still null 
and the   value returned by {\tt insert(s,x)} is 0, which means that
the element {\tt x} was not inserted. 

The last 
axiom can be read as: 
if 
the
set is neither null, %
 full nor empty and there is no node in the list with 
value {\tt x}, 
then, after execution, 
the set remains non-null
and non-empty, the value {\tt x} is now in the set, the
length is increased by 1, and the call to {\tt insert(s,x)} returns 1, which
denotes a successful insertion.
\end{example}

%% file: Sections/KFramework.tex

\K\  is a rewriting-based framework for engineering language semantics \cite{rosu_overview_2010}. Provided that the syntax and semantics of a programming language are formalized in the language of \K,\, 
the system automatically generates a parser, an interpreter, and formal analysis tools such as model checkers and deductive theorem provers.
Complete formal program semantics are currently available in \K\ 
for  Scheme, Java 1.4, JavaScript, Python, Verilog, and \C\ among
others 
\cite{rosu_overview_2010}.

A language definition in \K\ 
consists of three parts: the
BNF language syntax, 
the
structure of program configurations, and the semantic rules.  Program
configurations are represented in \K\ as potentially nested structures
of labeled {\em cells} (or containers) that represent the program
state.
Similarly to the classic operational semantics, program configuration cells 
include  a computation stack or continuation (named \kcompcell), one or more environments (\env, \heap), and a call stack (\stack) among others, and are represented as algebraic datatypes in \K. 

 The part of the \K\ program configuration structure for the \KernelC\ semantics that is relevant to this work is
$
\kallLarge{cfg}{
  \kall{k}{\operatorname{K}}
  \kall{env}{\operatorname{Map}}
  \kall{heap}{\operatorname{Map}}
}
$,
where the \env\ cell is a mapping of variable names to their memory
positions, the \heap\ cell binds the active memory positions to the
actual values (\ie{} it stores information about pointers and
data structures), and the \kcompcell\ cell represents a stack of
computations waiting to be run, with the left-most 
element
of the stack being the next computation to be undertaken.
\noindent For example, 
 the 
configuration

\vspace{-1ex}
{\footnotesize
\begin{equation}\label{eq:kconfiguration}
\kallLarge{cfg}{
  \kall{k}{\tv(int, 0)}
  \kall{env}{\mapsTo{{\tt x}}{\symba{x}}}
  \kall{heap}{\mapsTo{\symba{x}}{\tv(int, 5)}}
}
\end{equation}
}
\vspace{-2ex}

\noindent models the final state of a computation whose return value is the
integer 0 (stored in the \kcompcell\ cell, which contains the current
code to be run), while program variable \texttt{x} (stored in the
\env\ cell) has the integer value 5 (stored in the memory address given by
$\symba{x}$ in the \heap\ cell).
The symbol $\tv$ is a language construction aimed to encapsulate typed values.
Variables representing symbolic memory
addresses are written in sans-serif font preceded by the {\sf\&} symbol.

The semantic rules in \K\ state how configurations (terms) evolve throughout the
computation. A useful feature of \K\ is that 
<<rules only need to mention the minimum part of the configuration
that is relevant for their operation>>.

For symbolic reasoning, \K\ uses a particular class of first-order formulas
with equality (encoded as {boolean non--ground terms} 
with constraints over them). 
\ These formulas, called
\emph{patterns}, specify those 
configurations that match the
pattern algebraic structure and that satisfy its constraints. For
instance, the 
pattern

\vspace{-2ex}

{\footnotesize
\begin{align*}
\kallLarge{cfg}{
  \kall{k}{\tv(int, 0)} \\
  \kall{env}{\ellipses \mapsTo{{\tt x}}{\symba{x}}, \mapsTo{{\tt s}}{\symba{s}} \ellipses} \\
  \kall{heap}{\ellipses \mapsTo{\symba{s}}{(\mapsTo{{\tt size}}{\symbv{s.size}}, \mapsTo{{\tt capacity}}{\symbv{s.capacity}})} \ellipses}
}\\
\kallLarge{path-condition}{\symba{s} \neq {\sf NULL} \wedge \symbv{s.size} \geq \symbv{s.capacity}}
\end{align*}
}%
\vspace{-3ex}

\noindent specifies the configurations  satisfying that:%
\begin{inparaenum}[1)]
\item{the \kcompcell\ cell  only  contains the integer value 0;}
\item{in the \env\ cell, program variable \texttt{x} (in typographic font) is associated to the memory address \symba{x} 
while \texttt{s} is bound  to the pointer \symba{s}; and}
\item{in the \heap\ cell, the field {\tt size} of (the data structure
    pointed by) \symba{s} (resp.\ its {\tt capacity} field)
    contains the symbolic value\footnote{Symbolic values are preceded by a
      question mark.} \symbv{s.size} (resp.\
      \symbv{s.capacity}).%
  }
\end{inparaenum}
Additionally, \symba{s} is not null and the value of its {\tt size}
field is greater than or equal to its {\tt capacity} field.

Since patterns allow logical variables and constraints over them, by using patterns, the \K\ execution principle (which is based on term rewriting) becomes \emph{symbolic execution}.  
Unlike concrete execution where the path taken is determined by the
input, in symbolic execution the program can take any feasible path and each possible
  path  is associated to a \emph{path condition}, which represents the conditions that input values have to
  satisfy in order to follow that path. The path condition is formed by  constraints that are gathered along the path taken by the
 execution to reach the current program point, so 
each symbolic execution
path stands for many actual program runs (in fact, for exactly the set
of runs whose concrete values satisfy the logical constraints).

Symbolic execution in \K\ relies on an automated transformation of
\K\ configurations and \K\ rules into corresponding symbolic \K\
configurations (\ie\ patterns) and symbolic \K\ rules that capture all
required symbolic ingredients: symbolic values for data structure
fields and program variables; path conditions that constrain the
variables in cells; multiple branches when a condition is reached
during execution, etc \cite{arusoaie_symbolic_2015}. The transformed, symbolic rules define how
symbolic configurations are rewritten during computation. Roughly
speaking,
by symbolically executing a program statement, the configuration cells 
 are updated by mapping fields and variables to new symbolic values
 that are represented as symbolic expressions, while the path
 conditions (stored in a new \pathcondcell\ cell) are correspondingly
 updated at each branching point.

In \cite{AFV2015}, an 
inference procedure for \KernelC\  programs was defined using the \K\ symbolic
 execution infraestructure described above. In order to avoid the exponential blowup that is inherent to path enumeration, the symbolic procedure of \cite{AFV2015}
follows the standard approach of 
exploring loops up to a specified number of unfoldings. This ensures that symbolic execution ends for all explored
paths, thus delivering a finite (partial) representation of the program behavior [10].
In the following, given a   method call $m(\mathit{args})$ and an
initial path condition $\phi$, and assuming an 
unspecified 
 unrolling bound 
 for  loops, we denote by  \SE($m(\mathit{args})\{\phi\}$) the symbolic execution of method $m$ with input arguments $\mathit{args}$ as described
 in \cite{AFV2015}, which returns the set of leaves (patterns) of the symbolic execution tree for $m$
under the constraints given by $\phi$.
For any function $f$, by $f(\mathit{args})\{\phi\}$,
we represent the  \K\ pattern $\kall{cfg}{\kall{k}{f(\mathit{args})} 
\ellipses} \kall{path-condition}{\phi}$
that is built by inserting the call $f(\mathit{args})$ at the top of
the \kcompcell\ cell and by initializing the path condition cell with $\phi$.

%% file: Sections/Symbolic-Execution.tex
In this section,  we extend  \K's symbolic execution   machinery  with
lazy initialization techniques  and abstract subsumption checking 
in order to support the synthesis of contracts for methods that require refined loop finitization and
  \C\ pointer dereference and initialization.

\vspace{-1ex}

\paragraph{\bf\em Lazy initialization.}

Structured datatypes (\texttt{struct}) in \C\ are   aggregate types
that 
define   non-empty sets of sequentially allocated {\em
  member objects\footnote{An object in \C\ is a region of data storage
    in the execution environment.},} called fields, each of which has
a name and a type. 
In our symbolic setting,   pointer arithmetics and memory layout of  \C\  programs are abstracted by:%
\begin{inparaenum}[1)]
\item{operating with \emph{symbolic addresses} instead of concrete addresses, and} 
\item{mapping each structure object into a single element of the \heap\ cell that groups all object fields (and associated values).}
\end{inparaenum}
A specific field is then accessed by combining the identifiers of both the structure object and the field name.

A critical point in the symbolic execution of \C\ programs is the \emph{undefinedness} problem that occurs  when accessing uninitialized memory addresses. 
We adapt  the
lazy initialization approach of \cite{khurshid_generalized_2003} to
our setting as follows: when a symbolic address (or address expression) is
accessed for the first time, we force SE to initialize the memory object that is
located at the given address. 
This means
that the mapping in the \heap\ cell is updated by assigning a new
symbolic value (given by the very name of the
 symbolic address of the accessed field)
 that symbolically represents the assumptions 
made  on the dynamic data structure.
Actually, when symbolic execution accesses potentially
uninitialized 
memory positions, two  cases are considered:
the case in which the memory is 
initialized and it stores an
object of its respective type; and 
the case 
in which the memory stores a null pointer. 

To keep track of the constraints that are introduced by
  the lazy initialization, a new cell $\kall{init\text{-}heap}{}$ is added to the configuration that represents 
   the initialization  assumptions on the heap memory 
   at a given program point. In other words, at every leaf of the symbolic execution tree, the \initheap\ cell records the symbolic initial heap that leads to the given  final symbolic configuration.

\vspace{-1ex}
\paragraph{\bf\em Symbolic subsumption.}
Symbolic execution traditionally undergoes
non-termination problems in the presence of loops or recursion:   the exhaustive exploration of all   program paths
is unaffordable
 because the search space may be infinite and, consequently, the number
of symbolic execution paths may be unbounded. 
A classical solution (used in \cite{AFV2013,AFV2015}) is to establish a \emph{bound} to the depth of the symbolic execution tree by specifying the maximum number of unfoldings for each loop and recursive function. 
However, the completeness of the symbolic analysis is highly dependent on the chosen threshold,
and it is not  generally possible to 
ascertain the optimal number of iterations that \emph{subsume} all possible behaviors by   inspecting the source code.

The abstract subsumption approach of \cite{anand_symbolic_2008} 
determines the length of the symbolic execution paths in a dynamic way.
Intuitively, symbolic execution with abstract subsumption checking proceeds as standard symbolic execution, except that  before 
 entering a loop,  it is checked that the current (abstract) state has not already been explored; 
 otherwise, the execution of the loop stops.
Supporting this check does not require whole execution paths to be
recorded; only symbolic states that correspond to the
evaluation of loop guards need to be recorded.

An algorithm for symbolic  subsumption 
that naturally transfers 
to our framework is given in \cite{anand_symbolic_2008}. Let us augment  symbolic program configurations $C$
into \emph{program states} $S=\langle C,i\rangle$  by 
giving the configuration pattern $C$ a  program counter $i$ that corresponds to the
line number in the source code of  the subsequent instruction to be
executed, or the {\tt return} statement if the configuration $C$ is final. Also, let us represent the conjunction  of 
all constraints over the symbolic values of primitive-type
variables and structure fields expressed\footnote{By abuse, we assume a logical constraint representation $x_1= v_1 \wedge \ldots \wedge
   x_n= v_n$ of the  symbolic heap $\{x_1\mapsto v_1, \ldots,
   x_n\mapsto v_1\}$, where every $x_i$ references a field of a heap data object, whereas for the environment each $x_i$ refers to a primitive-type program variable. } 
   in the \env, \heap, and \pathcondcell\ cells of pattern $C$ in $S$, called \emph{state constraint}, by $\mathit{SC}(S)$.
By using the 
subsumption algorithm, we can decide  
state subsumption $S_2 \sqsubseteq  S_1$ by simply checking that:  
1) $S_1$ and $S_2$  have the same program counter; 2)  the symbolic heap in
$S_2$ is subsumed by the symbolic heap in $S_1$ (i.e.,\ the set of  all possible program heaps
whose concrete shape and values match the heap constraints in $S_1$ includes the set of all concrete program heaps
that satisfy the  constraints in $S_2$); 
and 
3) $\mathit{SC}(S_2) \Rightarrow \mathit{SC}(S_1)$.

\vspace{-1ex}

\paragraph{\bf\em Abstract subsumption.}

Symbolic execution with state subsumption 
is obviously not
guaranteed to terminate. In order to ensure 
termination and improve scalability of   symbolic execution, we 
enhance
symbolic state {subsumption} checking by means of abstract interpretation \cite{anand_symbolic_2008}. 
We  abstract both primitive domains and
heaps by using the  abstraction function $\alpha$ proposed in
\cite{anand_symbolic_2008}. The idea for heap abstraction is to apply a shape
transformation that collapses two or more nodes into a \emph{summary
  node}. Nodes can be collapsed when they are in a sequence and can only be
accessed by traversing all their predecessors 
(i.e., each node is only pointed to by its preceding node
in the sequence).

To make the visualization of symbolic heap abstraction easier, 
  we  also adapt to   our symbolic
 setting a classical graphical representation for heaps based on UML object diagrams,  
where  \emph{null} nodes are
rendered as ellipses,   uninitialized nodes are drawn as clouds, and
references 
are depicted as arrows. 

\begin{example}\label{ex:abstraction}
  Node $S_{27}$ of Figure~\ref{fig:abstractsubsumption}
illustrates shape abstraction for the
  given state. The circled nodes are abstracted into a summary
  node. Then, the first node of the list points to this new summary node and in turn the summary node points to
  the node referenced by \symba{end\_node}. Moreover, the valuation 
for the field {\tt value} of the summary node (identified by $e_3$) 
is $e_3 = \symbv{v$_0$} \vee e_3 = \symbv{v$_1$}$.

\end{example}

Given the symbolic state $S$, we define the abstraction  $S^\sharp=\alpha(S)$. Then, the abstract symbolic subsumption relation   $S_2\sqsubseteq^\sharp S_1$ is given by
 $S^\sharp_2\sqsubseteq  S^\sharp_1$.

\subsection{Symbolic execution with abstract subsumption}

The symbolic execution with abstract subsumption (and lazy
initialization) of a given method
$m$ with 
arguments $\mathit{args}$ and initial path condition $\phi$, written \SEAS($m(\mathit{args})\{\phi\}$), is defined as an approximation of the \SE\
mechanism of \cite{AFV2015} where, each time   a symbolic state $S_2$ is visited
that corresponds to a recursive call 
or loop guard 
evaluation
with the same program counter as a previously visited
state $S_1$, 
the  abstract
subsumption $S_2\sqsubseteq^\sharp S_1$ is checked; if the test succeeds, then the 
loop or recursive function stops, and the execution flow 
proceeds to the subsequent instruction.

\begin{figure}[ht!]
\vspace{-7ex}
\input{Insertables/Example-Subsumption.tex}
\vspace{-3ex}
\caption{Fragment of the abstract symbolic execution of  {\tt insert(s,x)}}
\label{fig:abstractsubsumption}
\vspace{-4ex}
\end{figure}
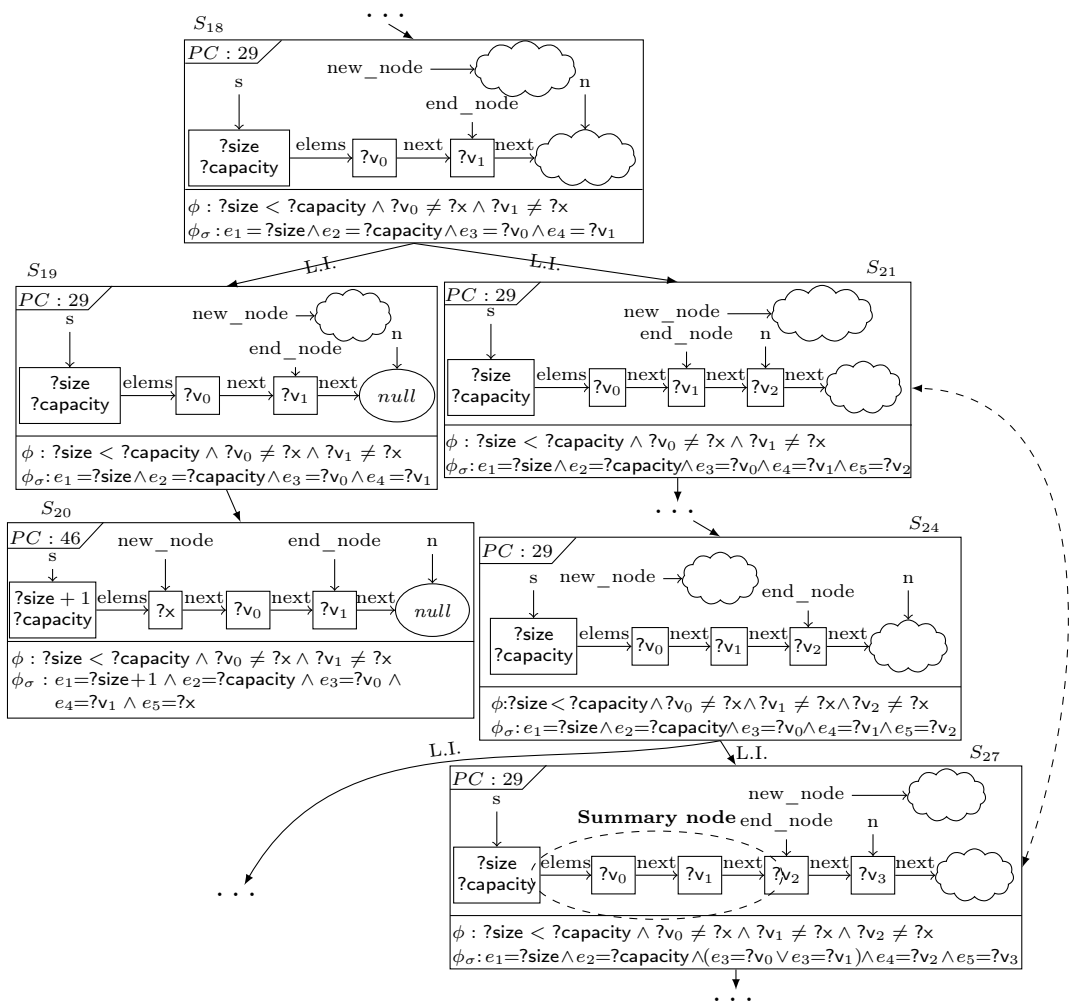

\begin{example}\label{ex:abstractsubsumption}
The uncontrolled symbolic execution \SE\ of the function {\tt insert(s,x)} from Example \ref{ex:program} 
  generates an infinite state space. 
 In contrast,   \SEAS\ 
terminates after three iterations of the loop. 
Figure~\ref{fig:abstractsubsumption} illustrates the fragment of the
symbolic execution tree for {\tt insert(s,x)} where the subsumption
between two abstract states is exposed.
  The  state ($S_{18}$) corresponds to the state
  where the loop guard 
 is to be checked for the third
  time. 
  This requires evaluating $n$, which points to an
  uninitialized node; hence, lazy initialization is applied. This
  results in two 
  children, $S_{19}$ and $S_{21}$, with the same program counter
  because the guard has not been evaluated yet.
  The left child $S_{19}$ corresponds to the case
  when the loop guard  is not satisfied and the loop is exited,
  whereas the   right child $S_{21}$ represents entering the loop
  iteration. 

Program counter 29 is reached again at state $S_{27}$ in the right branch after lazy
  initialization, and then the abstract subsumption check   $S_{27} \sqsubseteq^\sharp S_{21}$ succeeds. 

\end{example}

\newcommand*{\args}{\mathit{args}}

Let \SEAS$(f(\args)\{\phi\})$ return the set of final patterns obtained from the 
abstract symbolic execution
of the pattern $f(\args)\{\phi\}$ (\ie\ the leaves of the deployed abstract symbolic execution tree).
We assume appropriate abstractions are defined to ensure termination of \SEAS.
A new, (abstract subsumption) cell $\kall{aSubFlag}{}$  identifies with a $true$ value those  final  abstract configurations
ending any branch that was 
  folded  (at some intermediate configuration) by the application of abstract subsumption.
   This is used for
  the inference process  to distinguish the inferred axioms that are  
ensured to hold (because no approximation was done to extract them) from the  {\em plausible}, candidate axioms that are not demonstrably correct
because of the potential precision loss caused by the abstraction. 
Furthermore, in order to obtain the set of 
locations that may be affected by the execution of $f$ (the component
$\mathcal{L}$ of the contract), those locations have to be harvested   during 
symbolic execution. To this end, we add a new cell
$\kall{locations}{}$ to the symbolic engine of \K. Then, whenever a 
program location is overwritten, it 
is recorded 
 in the new cell \cellface{locations}. At the end of the symbolic
 execution, the program locations recorded for each final
 configuration in their respective \cellface{locations} cells are all
 joined by union to obtain a global set with every program location
 that is potentially modifiable by a call to the function $f$.
Thus, assignable locations are obtained as a by-product of
 the symbolic execution.

%% file: Insertables/Example-Subsumption.tex

\centerline{
{\scriptsize
\begin{tikzpicture}[scale=.8]\label{picture}
\draw (0,0) coordinate (root) {};
%
\draw (root)++(-4.4, -4) node [draw, rectangle, minimum width = 6.1cm, minimum height = 2.7cm, above right] (S18box) {};
\draw (S18box)++(-3.8,1.7) node [above right] (S18) {$S_{18}$};
%
\draw (S18box)++(-3.8,1.3) -- ++(1.2,0) -- ++(0.4,0.38);
\draw (S18box)++(-3.9,1.7) node [below right] (PC18) {$PC:29$};
%
\draw (S18box)++(-3.75,-0.75) node [draw, rectangle, above right] (s18) {$\begin{array}{c} \symbv{size} \\ \symbv{capacity} \end{array}$};
\node [draw, rectangle, right of = s18, node distance = 1.8cm, minimum height = 0.5cm] (v0s18) {$\symbv{v$_0$}$};
\node [draw, rectangle, right of = v0s18, node distance = 1.3cm, minimum height = 0.5cm] (v1s18) {$\symbv{v$_1$}$};
\node [draw, cloud, cloud puffs=11, cloud puff arc=120, aspect=2.3, inner ysep=0.2cm, right of = v1s18, node distance = 1.5 cm] (u1s18) {};
\draw (S18box)++(1.85,1.2) node [draw, cloud, cloud puffs=11, cloud puff arc=120, aspect=2.3, inner ysep=0.2cm] (u2s18) {};
%
\node [above of = s18] (sref18) {s};
\node [left of = u2s18, node distance = 2cm] (newnode18) {new\_node};
\node [above of = v1s18, node distance = 0.7cm] (endnode18) {end\_node};
\node [above of = u1s18] (n18) {n};
%
\draw [->] (sref18) -- (s18);
\draw [->] (s18) -- node [above] {elems} (v0s18);
\draw [->] (v0s18) -- node [above] {next} (v1s18);
\draw [->] (v1s18) -- node [above] {next} (u1s18);
\draw [->] (newnode18) -- (u2s18);
\draw [->] (endnode18) -- (v1s18);
\draw [->] (n18) -- (u1s18);
%
\draw (S18box)++(-3.8,-0.8) -- ++(7.6,0);
\draw (S18box)++(-3.95,-0.8) node [below right] (phi18) {$\begin{aligned} \phi : \symbv{size} < \symbv{capacity} \wedge \symbv{v$_0$} \neq \symbv{x} \wedge \symbv{v$_1$} \neq \symbv{x} \end{aligned}$};
\draw (S18box)++(-3.95,-1.2) node [below right] (val18) {$\begin{aligned} \phi_\sigma\!:\! e_1\! =\! \symbv{size}\! \wedge\! e_2\! =\! \symbv{capacity}\! \wedge\! e_3\! =\! \symbv{v$_0$}\! \wedge\! e_4\! =\! \symbv{v$_1$} \end{aligned}$};
%
\draw (root)++(-7.2, -8.1) node [draw, rectangle, minimum width = 5.6cm, minimum height = 2.7cm, above right] (S19box) {};
\draw (S19box)++(-3.45,1.7) node [above right] (S19) {$S_{19}$};
%
\draw (S19box)++(-3.5,1.3) -- ++(1.2,0) -- ++(0.4,0.395);
\draw (S19box)++(-3.6,1.7) node [below right] (PC19) {$PC:29$};
%
\draw (S19box)++(-3.45,-0.6) node [draw, rectangle, above right] (s19) {$\begin{array}{c} \symbv{size} \\ \symbv{capacity} \end{array}$};
\node [draw, rectangle, right of = s19, node distance = 1.7cm, minimum height = 0.5cm] (v0s19) {$\symbv{v$_0$}$};
\node [draw, rectangle, right of = v0s19, node distance = 1.3cm, minimum height = 0.5cm] (v1s19) {$\symbv{v$_1$}$};
\node [draw, ellipse, right of = v1s19, node distance = 1.35cm, minimum size = 0.7cm] (nulls19) {$null$};
\draw (S19box)++(2.1,1.2) node [draw, cloud, cloud puffs=11, cloud puff arc=120, aspect=1.7, inner ysep=0.2cm] (u2s19) {};
%
\node [above of = s19] (sref19) {s};
\node [left of = u2s19, node distance = 1.5cm] (newnode19) {new\_node};
\node [above of = v1s19, node distance = 0.6cm] (endnode19) {end\_node};
\node [above of = nulls19, node distance = 0.8cm] (n19) {n};
%
\draw [->] (sref19) -- (s19);
\draw [->] (s19) -- node [above] {elems} (v0s19);
\draw [->] (v0s19) -- node [above] {next} (v1s19);
\draw [->] (v1s19) -- node [above] {next} (nulls19);
\draw [->] (newnode19) -- (u2s19);
\draw [->] (endnode19) -- (v1s19);
\draw [->] (n19) -- (nulls19);
%
\draw (S19box)++(-3.5,-0.8) -- ++(7,0);
\draw (S19box)++(-3.6,-0.8) node [below right] (phi19) {$\begin{aligned} \phi : \symbv{size} < \symbv{capacity} \wedge \symbv{v$_0$} \neq \symbv{x} \wedge \symbv{v$_1$} \neq \symbv{x} \end{aligned}$};
\draw (S19box)++(-3.6,-1.2) node [below right] (val19) {$ \begin{aligned} \phi_\sigma\!\!:\! e_1 \!=\!\! \symbv{size} \!\wedge \! e_2 \!=\! \!\symbv{capacity} \!\wedge \! e_3 \! =\! \!\symbv{v$_0$} \!\wedge \! e_4 \! =\! \!\symbv{v$_1$} \end{aligned}$};
%
\draw (S19box)++(-3.65, -5.5) node [draw, rectangle, minimum width = 6.2cm, minimum height = 2.6cm, above right] (S20box) {};
\draw (S20box)++(-3.45,1.6) node [above right] (S20) {$S_{20}$};
%
\draw (S20box)++(-3.88,1.2) -- ++(1.2,0) -- ++(0.4,0.43);
\draw (S20box)++(-3.97,1.6) node [below right] (PC20) {$PC : 46$};
%
\draw (S20box)++(-3.85,-0.3) node [draw, rectangle, above right] (s20) {$\begin{array}{c}\! \!\symbv{size}+1 \!\! \\ \!\!\symbv{capacity}\!\! \end{array}$};
\node [draw, rectangle, right of = s20, node distance = 1.5cm, minimum height = 0.5cm] (xs20) {$\symbv{x}$};
\node [draw, rectangle, right of = xs20, node distance = 1.1cm, minimum height = 0.5cm] (v0s20) {$\symbv{v$_0$}$};
\node [draw, rectangle, right of = v0s20, node distance = 1.15cm, minimum height = 0.5cm] (v1s20) {$\symbv{v$_1$}$};
\node [draw, ellipse, right of = v1s20, node distance = 1.3cm ,
 minimum size = 0.7cm
] (nulls20) {$\mathit{null}$};
%
\node [above of = s20, node distance = 0.7cm] (sref20) {s};
\node [above of = xs20, node distance = 0.9cm] (newnode20) {new\_node};
\node [above of = v1s20, node distance = 0.9cm] (endnode20) {end\_node};
\node [above of = nulls20, node distance = 0.9cm] (n20) {n};
%
\draw [->] (sref20) -- (s20);
\draw [->] (s20) -- node [above] {elems} (xs20);
\draw [->] (xs20) -- node [above] {next} (v0s20);
\draw [->] (v0s20) -- node [above] {next} (v1s20);
\draw [->] (v1s20) -- node [above] {next} (nulls20);
\draw [->] (newnode20) -- (xs20);
\draw [->] (endnode20) -- (v1s20);
\draw [->] (n20) -- (nulls20);
%
\draw (S20box)++(-3.88,-0.35) -- ++(7.76,0);
\draw (S20box)++(-4,-0.35) node [below right] (phi20) {$\begin{aligned} \phi : \symbv{size} < \symbv{capacity} \wedge \symbv{v$_0$} \neq \symbv{x} \wedge \symbv{v$_1$} \neq \symbv{x} \end{aligned}$};
\draw (S20box)++(-4,-0.7) node [below right] (val20)
{$\begin{aligned} \phi_\sigma: {} & 
e_1\!\! =\!\! \symbv{size}\!+\!1 \wedge e_2\!\! =\!\!
    \symbv{capacity} \wedge e_3\!\! =\!\! \symbv{v$_0$} \wedge {} \\[-.75ex] & 
e_4 \!\!=\! \!\symbv{v$_1$} \wedge e_5\!\! =\!\! \symbv{x} \end{aligned}$};
%
\draw (S18box)++(0.5, -5.6) node [draw, rectangle, minimum width = 6.2cm, minimum height = 2.6cm, above right] (S21box) {};
\draw (S21box)++(3,1.6) node [above right] (S21) {$S_{21}$};
%
\draw (S21box)++(-3.88,1.25) -- ++(1.2,0) -- ++(0.4,0.38);
\draw (S21box)++(-4,1.65) node [below right] (PC21) {$PC : 29$};
%
\draw (S21box)++(-3.83,-0.6) node [draw, rectangle, above right] (s21) {$\begin{array}{c} \!\!\symbv{size}\!\! \\ \!\! \symbv{capacity}\!\! \end{array}$};
\node [draw, rectangle, right of = s21, node distance = 1.55cm, minimum height = 0.5cm] (v0s21) {$\!\symbv{v$_0$}\!$};
\node [draw, rectangle, right of = v0s21, node distance = 1.05cm, minimum height = 0.5cm] (v1s21) {$\!\symbv{v$_1$}\!$};
\node [draw, rectangle, right of = v1s21, node distance = 1.05cm, minimum height = 0.5cm] (v2s21) {$\!\symbv{v$_2$}\!$};
\node [draw, cloud, cloud puffs=11, cloud puff arc=120, aspect=1.7, inner ysep=0.2cm, right of = v2s21, node distance = 1.3 cm] (u1s21) {};
\draw (S21box)++(2.4,1.1) node [draw, cloud, cloud puffs=11, cloud puff arc=120, aspect=2.3, inner ysep=0.2cm] (u2s21) {};
%
\node [above of = s21] (sref21) {s};
\node [left of = u2s21, node distance = 2cm] (newnode21) {new\_node};
\node [above of = v1s21, node distance = 0.7cm] (endnode21) {end\_node};
\node [above of = v2s21, node distance = 0.7cm] (n21) {n};
%
\draw [->] (sref21) -- (s21);
\draw [->] (s21) -- node [above] {elems} (v0s21);
\draw [->] (v0s21) -- node [above] {next} (v1s21);
\draw [->] (v1s21) -- node [above] {next} (v2s21);
\draw [->] (v2s21) -- node [above] {next} (u1s21);
\draw [->] (newnode21) -- (u2s21);
\draw [->] (endnode21) -- (v1s21);
\draw [->] (n21) -- (v2s21);
%
\draw (S21box)++(-3.87,-0.75) -- ++(7.76,0);
\draw (S21box)++(-4.05,-0.75) node [below right] (phi21) {$\begin{aligned} \phi: \symbv{size} < \symbv{capacity} \wedge \symbv{v$_0$} \neq \symbv{x} \wedge \symbv{v$_1$} \neq \symbv{x} \end{aligned}$};
\draw (S21box)++(-4.05,-1.15) node [below right] (val21)
{$\begin{aligned} \phi_\sigma\!\!:\! 
e_1 \!\!=\! \!\symbv{size} \!\wedge\! e_2\! \!=\!\!
    \symbv{capacity} \!\!\wedge\! e_3\!\! =\!\! \symbv{v$_0$}\!\! \wedge\! {} 
e_4\!\! =\!\! \symbv{v$_1$}\!\! \wedge\! e_5\!\! =\!\! \symbv{v$_2$} \end{aligned}$};
%
\draw (S21box)++(-3.3, -6) node [draw, rectangle, minimum width = 6.4cm, minimum height = 2.7cm, above right] (S24box) {};
\draw (S24box)++(3,1.65) node [above right] (S24) {$S_{24}$};
%
\draw (S24box)++(-4,1.25) -- ++(1.2,0) -- ++(0.4,0.43);
\draw (S24box)++(-4.1,1.7) node [below right] (PC24) {$PC : 29$};
%
\draw (S24box)++(-3.83,-0.6) node [draw, rectangle, above right] (s24) {$\begin{array}{c} \!\!\symbv{size} \!\!\\\!\! \symbv{capacity}\!\! \end{array}$};
\node [draw, rectangle, right of = s24, node distance = 1.55cm, minimum height = 0.5cm] (v0s24) {$\!\symbv{v$_0$}\!$};
\node [draw, rectangle, right of = v0s24, node distance = 1.05cm, minimum height = 0.5cm] (v1s24) {$\!\symbv{v$_1$}\!$};
\node [draw, rectangle, right of = v1s24, node distance = 1.05cm, minimum height = 0.5cm] (v2s24) {$\!\symbv{v$_2$}\!$};
\node [draw, cloud, cloud puffs=11, cloud puff arc=120, aspect=1.75, inner ysep=0.2cm, right of = v2s24, node distance = 1.33 cm] (u1s24) {};
\draw (S24box)++(0,1) node [draw, cloud, cloud puffs=11, cloud puff arc=120, aspect=1.7, inner ysep=0.2cm] (u2s24) {};
%
\node [above of = s24, node distance = 0.9cm] (sref24) {s};
\node [left of = u2s24, node distance = 1.5cm] (newnode24) {new\_node};
\node [above of = v2s24, node distance = 0.7cm] (endnode24) {end\_node};
\node [above of = u1s24, node distance = 0.9cm] (n24) {n};
%
\draw [->] (sref24) -- (s24);
\draw [->] (s24) -- node [above] {elems} (v0s24);
\draw [->] (v0s24) -- node [above] {next} (v1s24);
\draw [->] (v1s24) -- node [above] {next} (v2s24);
\draw [->] (v2s24) -- node [above] {next} (u1s24);
\draw [->] (newnode24) -- (u2s24);
\draw [->] (endnode24) -- (v2s24);
\draw [->] (n24) -- (u1s24);
%
\draw (S24box)++(-4,-0.8) -- ++(8,0);
\draw (S24box)++(-4,-0.8) node [below right] (phi24) {$\begin{aligned} \phi\!\! :\!\! \symbv{size}\! <\! \symbv{capacity}\! \wedge\! \symbv{v$_0$} \neq \symbv{x}\! \wedge\! \symbv{v$_1$} \neq \symbv{x} \!\wedge\! \symbv{v$_2$} \neq \symbv{x} \end{aligned}$};
\draw (S24box)++(-4,-1.2) node [below right] (val24)
{
$\begin{aligned} \phi_\sigma\!\!:\! 
e_1 \!\!=\! \!\symbv{size} \!\wedge\! e_2\! \!=\!\!
    \symbv{capacity} \!\!\wedge\! e_3\!\! =\!\! \symbv{v$_0$}\!\! \wedge\! {} 
e_4\!\! =\!\! \symbv{v$_1$}\!\! \wedge\! e_5\!\! =\!\! \symbv{v$_2$} \end{aligned}$
};
%
\draw (S24box)++(-4.5, -5.5) node [draw, rectangle, minimum width = 7.6cm, minimum height = 2.7cm, above right] (S27box) {};
\draw (S27box)++(3.75,1.65) node [above right] (S27) {$S_{27}$};
%
\draw (S27box)++(-4.75,1.3) -- ++(1.2,0) -- ++(0.4,0.4);
\draw (S27box)++(-4.85,1.7) node [below right] (PC27) {$PC : 29$};
%
\draw (S27box)++(-4.7,-0.6) node [draw, rectangle, above right] (s27) {$\begin{array}{c} \!\!\symbv{size}\!\! \\\!\! \symbv{capacity}\!\! \end{array}$};
\node [draw, rectangle, right of = s27, node distance = 1.55cm, minimum height = 0.5cm] (v0s27) {$\symbv{v$_0$}$};
\node [draw, rectangle, right of = v0s27, node distance = 1.15cm, minimum height = 0.5cm] (v1s27) {$\symbv{v$_1$}$};
\node [draw, rectangle, right of = v1s27, node distance = 1.15cm, minimum height = 0.5cm] (v2s27) {$\symbv{v$_2$}$};
\node [draw, rectangle, right of = v2s27, node distance = 1.15cm, minimum height = 0.5cm] (v3s27) {$\symbv{v$_3$}$};
\node [draw, cloud, cloud puffs=11, cloud puff arc=120, aspect=1.75, inner ysep=0.2cm, right of = v3s27, node distance = 1.35 cm] (u1s27) {};
\draw (S27box)++(3.5,1.2) node [draw, cloud, cloud puffs=11, cloud puff arc=120, aspect=1.75, inner ysep=0.2cm] (u2s27) {};
%
\node [above of = s27] (sref27) {s};
\node [left of = u2s27, node distance = 2cm] (newnode27) {new\_node};
\node [above of = v2s27, node distance = 0.7cm] (endnode27) {end\_node};
\node [above of = v3s27, node distance = 0.7cm] (n27) {n};
%
\draw [->] (sref27) -- (s27);
\draw [->] (s27) -- node [above] {elems} (v0s27);
\draw [->] (v0s27) -- node [above] {next} (v1s27);
\draw [->] (v1s27) -- node [above] {next} (v2s27);
\draw [->] (v2s27) -- node [above] {next} (v3s27);
\draw [->] (v3s27) -- node [above] {next} (u1s27);
\draw [->] (newnode27) -- (u2s27);
\draw [->] (endnode27) -- (v2s27);
\draw [->] (n27) -- (v3s27);
%
\draw (S27box)++(-4.75,-0.8) -- ++(9.5,0);
\draw (S27box)++(-4.9,-0.8) node [below right] (phi27) {$\begin{aligned} \phi: \symbv{size} < \symbv{capacity} \wedge \symbv{v$_0$} \neq \symbv{x} \wedge \symbv{v$_1$} \neq \symbv{x} \wedge \symbv{v$_2$} \neq \symbv{x} \end{aligned}$};
\draw (S27box)++(-4.9,-1.2) node [below right] (val27) {$\begin{aligned}
    \phi_\sigma\!\!:\! {} & e_1\!\! =\!\! \symbv{size}\! \wedge\! e_2\!\! =\!\!
    \symbv{capacity} \!\wedge\!\! (e_3\! \!=\!\! \symbv{v$_0$} \!\vee\! e_3 \!\!=\!\! \symbv{v$_1$})\!
    \!\wedge
\!e_4 \!\!=\!\! \symbv{v$_2$}\! \wedge\! e_5 \!\!=\!\! \symbv{v$_3$} \end{aligned}$};
%
%
\draw (root)++(-1, -0.2) node (initialellipses) {\Large \ldots};
\node [below of = S21box, node distance = 1.75cm] (ellipses2124) {\Large \ldots};
\draw (S24box)++(-8,-4.25) node (ellipses24down) {\Large \ldots};
\node [below of = S27box, node distance = 1.75cm] (ellipses27down) {\Large \ldots};
\draw [->, >=latex] (initialellipses.south) -- (S18box.north);
\draw [->, >=latex] (S18box.south) -- node [sloped] {L.I.} (S19box.north);
\draw [->, >=latex] (S19box.south) -- (S20box.north);
\draw [->, >=latex] (S18box.south) -- node [sloped] {L.I.} (S21box.north);
\draw [->, >=latex] (S21box.south) -- (ellipses2124);
\draw [->, >=latex] (ellipses2124) -- (S24box.north);
\draw [->, >=latex] (S24box.south) to [out=190, in=60] node [sloped, above] {L.I.} (ellipses24down);
\draw [->, >=latex] (S24box.south) -- node [right] {L.I.} (S27box.north);
\draw [->, >=latex] (S27box.south) -- (ellipses27down);
\draw [<->, >=latex, dashed] (S27box.east) to [out=70, in=358] (S21box);
%
\node [draw, ellipse, fit=(v0s27) (v1s27), dashed] (summarynode) {};
\node [above of = summarynode, node distance = 0.75cm] () {\textbf{Summary node}}; 
\end{tikzpicture}
}
}

%% file: Sections/Inference.tex
{Let us introduce the basic notions that we use in our formalization. Given an input program $P$, 
we distinguish the set of observers $\mathcal{O}$ and the set of modifiers $\mathcal{M}$ in $P$. }
A function can be considered to be an \emph{observer} if it explicitly returns a value, whereas any method can be considered to be a \emph{modifier}. Thus, the set $\mathcal{O} \cap \mathcal{M}$ is generally non-empty.

\newcommand{\ol}[1]{\overline{#1}}  
\vspace{-1ex}
\begin{algorithm}[h]

\renewcommand{\baselinestretch}{0.97}
\caption{Specification Inference}
\label{alg:specinference}
\begin{algorithmic}[1]
{
\REQUIRE $m \in \mathcal{M}:$ a modifier function with arity $n$
\ENSURE $\mathit{contract}:$ a specification contract for $m$

\STATE $\mathit{root}:=m(\ol{a_n})$;
\STATE ${\cal F}:=$\SEAS$(\mathit{root}\{\texttt{true}\})$;
\STATE $\mathit{P} := \mathit{false}$;
$\mathit{Q} := \mathit{true}$;
$\mathit{Q^\sharp} := \mathit{true}$;
$\mathit{\cal L} := \emptyset$;
\FORALL {$F \in {\cal F}$, with $F = \kall{cfg}{\kall{k}{v} \kall{init-heap}{\varphi} \ellipses}
 						\kall{path-condition}{\phi} \kall{aSubFlag}{\sharp} \kall{locations}{\mathit{L}}$ 
}
	\STATE $p := \explain{
						I,
						\ol{a_n}}$, where $I=\kall{cfg}{\kall{k}{\mathit{root}} \kall{heap}{\varphi} \ellipses}					\kall{path-condition}{\phi}$;
	\STATE $\mathit{q} := \mathit{explain}(F, \ol{a_n}) \wedge (\mathit{ret} = v)$;
	\STATE $ \mathit{P} := \mathit{P} \vee p$;
	\STATE $ax := (p \Rightarrow q)$;
	\STATE \textbf{if} $\sharp$ \textbf{then} 
             {$\mathit{Q} := \mathit{Q} \wedge ax$;} \textbf{else}
              {$\mathit{Q^\sharp} := \mathit{Q^\sharp} \wedge ax$;}
	\STATE ${\cal L} := {\cal L} \cup \{\mathit{L}\}$;
\ENDFOR
\RETURN $<\mathit{P},\mathit{refine}(\mathit{Q,Q^\sharp}),{\cal L}>$; 
}\end{algorithmic}

\end{algorithm}
\vspace{-1ex}

Our specification inference methodology is formalized in Algorithm
\ref{alg:specinference}.
Let $\ol{a_n}$ denote the list of fresh symbolic variables $a_1, \ldots, a_n$.
First, the \emph{modifier} method of interest
$m$ is symbolically executed with   argument list 
$\ol{a_n}$ and empty path constraint \texttt{true}, and the
set $\cal F$ of final 
configurations is retrieved from the leaves of the abstract symbolic
execution tree. For each final
configuration 
the corresponding path
condition 
$\phi$ is simplified by calling the automated theorem prover Z3.

After initializing the contract components (Line~3), we proceed to compute one axiom for each 
(abstract) symbolic configuration 
$F$ in $\cal F$.
First, the 
premise $p$ of the axiom $p \Rightarrow q$ 
is computed (Line~5) by 
means of the function
$\explain{I,as}$ originally proposed in \cite{AFV2013}. 
This function 
receives as argument the pattern $I$, which expresses the initial symbolic configuration leading to $F$ in the  execution tree  for $m$ (\ie\ a variant of the  initial configuration  
for $m(\ol{a_n}$) that is obtained by assuming the  constraints $\varphi$ and $\phi$ 
in the corresponding \initheap\ and \pathcondcell\ cells of $I$).
Roughly speaking, 
by means of a conjunction of equations $\explain{I,\ol{a_n}}$
describes  what can be observed when running (under the constraints given by $I$) the observer functions $o\in\mathcal{O}$  over appropriate symbolic variables   from   $\ol{a_n}$. Each delivered equation is formed by equating each  observer call 
 to the (symbolic) value that the call returns.
We require $o$ to compute the same symbolic values at the end of all its symbolic execution branches
in order to distill a (partial) observational abstraction or
explanation for 
a given configuration in terms of  $o$.  

The 
consequent $q$ of the axiom is the conjunction 
of  
$ret = v$, which specifies  the return
value $v$ of the method $m$ as recorded in the \kcompcell\ cell of $F$, and the   equations
given by $\explain{F, \ol{a_n}}$, which  in terms of the observers characterizes the final
pattern $F$ of the given branch.
Note that the return value $v$ could be either 
{\sf uninit}
or
an initialized typed value that represents the return value for $m$ under the conditions given by $\phi$. 

It is important to note that, in the axioms, the different equations
in the antecedent (resp. consequent) of every implication formula are
assumed to be
run independently of each other under the same initial
configuration. This is achieved by the $\explain{}$ algorithm by using
the same initial state when considering the different observer
functions to explain $I$ and $F$. This avoids making any assumptions
about function purity or side-effects.
Depending on the boolean value of the abstract subsumption
flag $\sharp$ in $F$ (line~9), the synthesized axiom $ax$ is directly added 
 to the postcondition ${\mathit Q}$ (when $\sharp$ is $\mathit{false}$) or  to the conjunction ${\mathit Q}^\sharp$ (when $\sharp$ is true) that collects all candidate axioms extracted from branches that contain at least one node that was folded by abstract subsumption. Note that, due to the under-approximation introduced by  abstract subsumption \cite{anand_symbolic_2008}, there may be some behaviors
(real trace fragments) beyond the abstract folded states that are not captured by the 
deployed  symbolic abstract traces.
Therefore, axioms 
in ${\mathit Q^\sharp}$ could have spurious instances and must be double-checked.
We apply a post-processing refinement    
$\mathit{refine}({\mathit Q},{\mathit Q^\sharp})$ 
which tries to 
build specialized  (demonstrably correct) 
instances of the axioms in ${\mathit Q^\sharp}$ that can be added to ${\mathit Q}$,
  while getting rid of any ${\mathit Q^\sharp}$ axioms that remain
  overly general (i.e.,\ that can have both true and false instances).
  A further subsumption checking over the resulting set of axioms is included  in the refinement post-processing that 
purges the augmented ${\mathit Q}$ from less general axioms.

When Algorithm \ref{alg:specinference} terminates,  the generated contract is $<\mathit{P},\mathit{Q},{\cal L}>$ where the   method precondition  $\mathit{P}$ is the disjunction of all axiom premises, the method postcondition is given by $\mathit{refine}(\mathit{Q},\mathit{Q}^\sharp)$, and ${\cal L}$ records
all program locations that are (potentially) modifiable by $m$. Note that we do not need to specialize the disjunction $\mathit{P}$
according to the final refined postcondition $\mathit{Q}$ because correctness of the contract is ensured by the   specialized axiom guards of  $\mathit{Q}$.

We note that lazy initialization is not applied when   symbolically executing the observer functions. This is because we want  to start from an initial configuration whose  dynamic memory satisfies (or is given by) $\varphi$, and if any uninitialized addresses are expanded by lazy initialization, such an  initial configuration (and thus the target of the observation) would be altered. 
This implies that some final patterns in the symbolic execution trees
for the given observer may contain {\sf uninit} return values,
meaning that we know nothing regarding the dynamic memory from
that point on. When this occurs (for all branches), the $\explain{}$
algorithm generates a conjunct where the  observer call is equated  to a fresh symbolic value.

Let us compute a specification for the {\tt insert} modifier function
of Example~\ref{ex:program} by applying
Algorithm~\ref{alg:specinference}.
\begin{example}\label{ex:mainprocess}
We first compute  \SEAS$({\tt insert}(\symba{s}, \symbv{x})\{{\tt true}\})$ with \symba{s} 
being a symbolic address with initial value {\sf uninit} and with
\symbv{x} being a symbolic integer value. 
Since there are no constraints in the initial symbolic configuration, the execution covers all possible initial concrete configurations. 
Then, the abstract symbolic execution computes ten final configurations. The
following one represents the final state for the path where the
{\tt while} loop stops due to abstract subsumption between the states
associated to two consecutive iterations (Nodes $S_{18}$ and $S_{27}$
of Example~\ref{ex:abstractsubsumption}):

{\scriptsize
\[
\begin{array}{c}
\kall{k}{\tv(int, 1)} \\
{\scaleleftright[1ex]{<}
{\begin{array}{c}
 	\mapsTo{{\tt s}}{\symba{s}},
 	\mapsTo{{\tt x}}{\symbv{x}},
 	\mapsTo{{\tt new\_node}}{\symba{new\_node}},
 	\\
	\mapsTo{{\tt end\_node}}{\symba{s.elems.next.next}},
	\mapsTo{{\tt n}}{\symba{s.elems.next.next.next}}
\end{array}}{>}}
_\cellface{env}\\
{\scaleleftright[1ex]{<}
{\begin{array}{c}
	\mapsTo{\symba{s}}{(
		\mapsTo{{\tt capacity}}{\symbv{s.capacity}},
		\mapsTo{{\tt size}}{\symbv{s.size + 1}},
		\mapsTo{{\tt elems}}{\symba{new\_node}}
	)},
	\\
	\mapsTo{\symba{new\_node}}{(
		\mapsTo{{\tt value}}{\symbv{x}},
		\mapsTo{{\tt next}}{\symba{s.elems}}
	)}, 	\mapsTo{\symba{x}}{\symbv{x}}
	\\
	\mapsTo{\symba{s.elems}}{(
		\mapsTo{{\tt value}}{\symbv{v$_0$}},
		\mapsTo{{\tt next}}{\symba{s.elems.next}}
	)},
	\\
\ldots
	\\
	\mapsTo{\symba{s.elems.next.next.next}}{(
		\mapsTo{{\tt value}}{\symbv{v$_3$}},
		\mapsTo{{\tt next}}{\textsf{uninit}}
	)}
\end{array}}{>}}
_\cellface{heap}
\\
{\scaleleftright[1ex]{<}
{\begin{array}{c}
	\mapsTo{\symba{s}}{(
		\mapsTo{{\tt capacity}}{\symbv{s.capacity}},
		\mapsTo{{\tt size}}{\symbv{s.size}},
		\mapsTo{{\tt elems}}{\symba{s.elems}}
	)},
	\\
	\mapsTo{\symba{s.elems}}{(
		\mapsTo{{\tt value}}{\symbv{v$_0$}},
		\mapsTo{{\tt next}}{\symba{s.elems.next}}
	)},
	\\
\ldots\\
	\mapsTo{\symba{s.elems.next.next.next}}{(
		\mapsTo{{\tt value}}{\symbv{v$_3$}},
		\mapsTo{{\tt next}}{\textsf{uninit}}
	)}
\end{array}}{>}}
_\cellface{init-heap}
\\
\kall{path-condition}{\symbv{s.size} < \symbv{s.capacity} \wedge \symbv{v$_0$} \neq \symbv{x} \wedge \symbv{v$_1$} \neq \symbv{x} \wedge \symbv{v$_2$} \neq \symbv{x}}
\end{array}
\]
}

 \vspace{-1ex}
Roughly speaking, the execution of this path corresponds to the case
when the element {\tt x} (with symbolic value \symbv{x}) is
effectively inserted in a non-empty 
list that contains three elements. 
The return value  (\kcompcell{} cell) of the call ${\tt
  insert}(\symba{s}, \symbv{x})$ is the integer 1 (standing for
success); the symbolic (initial) value \symbv{s.size} 
of the field {\tt size} of {\tt s} is increased by 1 and now the field
{\tt elems} of {\tt s} points to an object \symba{new\_node} with {\tt
  value} \symbv{x} as the first node of the set. For the sake of
simplicity, we omit any cell components   that  are irrelevant
for comprehension.

As a side effect of applying abstract subsumption to stop the {\tt while} loop, 
the node pointed by the field {\tt next} of the last object node 
is not null but {\sf uninit}. 
 This implies a loss of precision: the symbolic heap is matched  by any concrete heap whose first node contains  the value
  \symbv{x} and is followed by 3 or more nodes.
\end{example}

The algorithm computes the
explanation for the corresponding initial and final state of each of
those ten configurations.
Let us illustrate one of the cases. 
\begin{example}
In order to explain the final pattern $F$ of Example~\ref{ex:mainprocess}, the function $\mathit{explain}$ considers the universe of observer calls, which include
the call {\tt contains(\symba{s},\symbv{x})}. 
The symbolic execution of {\tt contains(\symba{s},\symbv{x})} 
under the constraints given by the \heap\ and
\pathcondcell\ cells of $F$ 
results in a single-branch tree with return
value 1; hence, the equation 
 {\tt contains(s,x)=1} is added as part of the equational explanation of $F$.
\end{example}

\begin{example}\label{ex:initialpattern}
  In order to explain the corresponding initial pattern $I$, we symbolically execute  the observer {\tt
    contains(\symba{s},\symbv{x})} under the constraints given by $I$ (\ie\ the \initheap\ and \pathcondcell\ cells of $F$); and  since  no element  with   value \symbv{x} is found  in the  set \symba{s}, 
  the list
  is traversed until the \textsf{uninit} node is reached.
Hence,
the equation {\tt contains(s,x)=}$\_v$ is generated, with \_$v$ being a
symbolic  value  that stands for any possible value that the function
may return (either 0 or 1 in this example).  
\end{example}

Finally, let us illustrate how the refinement process $\mathit{refine}({\mathit Q},{\mathit Q}^\sharp)$ for method $m$ works.
Roughly speaking, for each candidate axiom $p\Rightarrow q$ in
  ${\mathit Q}^\sharp$, we first randomly generate  test cases 
(initial configurations) that satisfy the axiom antecedent $p$, then
we run the modifier method $m$ on  the  initial configurations, and
finally we check
  whether or not the consequent  $q$ is satisfied after the method
  execution. Refuted candidate axioms  are not automatically removed:
  a counterexample-guided,  specialization post-process {defined
    in \cite{AFV2013}} is attempted  first.
It uses  the concrete values refuting the axiom (or more precisely the deployed symbolic execution branches resulting from fixing  those values on the initial configuration), as counterexample behaviors to   be excluded from the symbolic execution tree.
Then,  by iteratively repeating the inference process on the reduced
tree,  new   axioms  that are either eventually correct (and then
added to ${\mathit Q}$) or can be further specialized are distilled. Note that this process is guaranteed
to terminate since  the size of the tree is reduced at each iteration.

\begin{example}
After the for {\tt loop} of Algorithm~\ref{alg:specinference}, 
  one axiom for each of the (10) final patterns 
is synthesized. After
  removing duplicates, 7 axioms are kept (see Figure~\ref{fig:axioms}), together
with one candidate axiom (labelled as {\tt C1}) that derives  from the final configuration discussed in
Example~\ref{ex:mainprocess}. 
  
\begin{figure}[ht!]
\vspace{-4ex}
\renewcommand{\baselinestretch}{0.97}

{\scriptsize \SpecActualSubsumption}
\vspace{-4ex}
\caption{Set of axioms and candidates for
  Example~\ref{ex:mainprocess}.\label{fig:axioms}}
\vspace{-4ex}
\end{figure}

The refinement process is then triggered over {\tt C1} 
to check if it can first be falsified and then
refined.
Given the binary domain 0/1 of the {\tt contains(s,x)} function, the
axiom is straightforwardly falsified (e.g.,\ by the test case where {\tt s} is
a non-full set containing a single element
with value 5, and {\tt x} is 5).
The final state 
does not satisfy 
the postcondition  of axiom {\tt C1}  because, since the set   {\tt s} already contained the   desired element,  
the modifier {\tt insert(s,x)} 
does not return 1 
and the length does not increase after the execution.

Since the axiom has been falsified (with {\tt contains(s,x)=1}), 
now the refinement process is run
with 
{\tt $\_i1 \mapsto 0$} and the  
last 
(specialized and correct) axiom is obtained: 

\vspace{-2ex}
{\scriptsize \[
{\tt A8} \left(
\begin{array}{l}
\mathtt{isnull(s)}=\mathtt{0} \; \wedge 
\mathtt{isempty(s)}=\mathtt{0} \; \wedge \\
\mathtt{isfull(s)}=\mathtt{0} \; \wedge 
\mathtt{contains(s,x)}=\mathtt{0} \; \wedge \\
\mathtt{length(s)}=\_i1
\end{array}
\right)
 \Rightarrow
\left(
\begin{array}{l}
\mathtt{isnull(s')}=\mathtt{0} \; \wedge 
\mathtt{isempty(s')}=\mathtt{0} \; \wedge \\
\mathtt{contains(s',x)}=\mathtt{1} \; \wedge 
\mathtt{length(s')}=\_i1+1 \; \wedge \\
\mathtt{ret}=\mathtt{1}
\end{array}
\right)
\]
\vspace{-3ex}}

Note that the new axiom subsumes the  fifth, sixth, and
seventh axioms of the previous specification; hence, they are removed. 
After the refinement, the contract postcondition returned for {\tt
  insert(s,x)} contains five axioms, specifically the axioms {\tt A1-4} and {\tt A8}. 
  
As for the last element of the contract, the set of assignable program locations $\mathcal{L}$ is obtained as the union of the  location sets 
that are recorded in the $\kall{locations}{}$ cells of the final
symbolic execution states, which is 
{\footnotesize $\mathcal{L} = \{{\tt s}, {\tt end\_node}, {\tt n}, {\tt new\_node}, {\tt new\_node \mapsto value},$\\ ${\tt new\_node \mapsto next}, 
{\tt s \mapsto elems}, {\tt s \mapsto size}\}$}.
\end{example}

%% file: Sections/RelatedConc.tex
The wide interest in formal specifications as helpers for other analysis, validation, and verification tools have resulted in numerous approaches for (semi-)automatically computing different kinds of 
specifications that
can take  the form of contracts, snippets, summaries, 
 properties, process models, rules, graphs, automata, interfaces, or component abstractions. 

Let us briefly discuss those strands of research that have influenced our work the most.
A  detailed description of the related literature
can be found  in \cite{wei_inferring_2011,AFV2013,Cousot2013}. 
Our axiomatic representation is inspired by \cite{tillmann_discovering_2006}, which relies on a model checker for symbolic execution and generates either Spec\# specifications or parameterized unit tests. In contrast to \cite{tillmann_discovering_2006}, we take advantage of \K\ symbolic capabilities to generate simpler and more accurate formulas that avoid reasoning with the global heap because the different pieces of the heap that are reachable from the function argument addresses are kept separate. 
Unlike our symbolic approach, \QuickSpec\ \cite{claessen_quickspec_2010}, Daikon \cite{ernst_daikon_2007}, and 
the algebraic specification discovery tool of Henkel and Diwan \cite{henkel_discovering_2003} 
detect program 
assertions   by extensive testing.
Whereas Daikon discovers invariants that hold at existing program points, \QuickSpec\ discovers equations between arbitrary terms (laws) that are constructed by using an API. This is similar 
to the approach of
 Henkel and Diwan \cite{henkel_discovering_2003}, which  generalizes  the results of  running tests on Java class interfaces as an algebraic specification.  By combining the concrete execution of actual test cases with a simultaneous symbolic execution of the same tests, DySy determines program properties    that generalize the observed  behaviors \cite{csallner_dysy_2008}.  Starting from simple, partial contracts previously written by the programmer, rich post-conditions involving quantification are defined in \cite{wei_inferring_2011} by using random testing.
Other approaches to software specification discovery    based on abstract interpretation   are
\cite{taghdiri_inferring_2006,bacci_automatic_2012,Cousot2013}, while   
  \cite{whaley_automatic_2002,giannakopoulou_interface_2009} use inductive matching learning.

Since our approach is generic and not tied to the \K\ semantics specification of \KernelC, we expect that the methodology developed in this work can be easily extended to other languages for which a \K\ semantics is given. 
Moreover, the correctness of the delivered specifications can be automatically ensured by using the existing \K\ formal tools.

 We have developed a prototype implementation of the extended \K\ symbolic machinery and contract
inference algorithm described in the previous sections (available at
\url{http://safe-tools.dsic.upv.es/kindspec2}), and we have used it to mechanize our runnning example.
The abstraction component is not fully integrated within the system, yet it can be used
by manually fixing the abstract domain for the program at hand. 
Our preliminary results are promising since they show that general correct axioms can be inferred, leading to a more compact, clear, 
and complete specification.
The contracts   generated by our tool can be easily translated to   richer (but also heavier) notations  for behavioural interface  \C\ specifications such as ACSL 
 or to  the native syntax of some SMT solvers, which is planned as future work.

%% file: small-bbl2.bbl

%% file: Sections/Appendix.tex
\section{Full example code}\label{apndx:full_program}
\begin{figure}[ht!]
\lstset{
language=C,
basicstyle={\ttfamily\scriptsize},
breaklines=true,
numbers=left,
stepnumber=1,
tabsize=1,
columns=[l|l]flexible
}
\setlength\columnsep{15pt}
\begin{multicols}{2}
\begin{lstlisting}
#include <stdlib.h>

struct lnode{
	int value;
	struct lnode *next;
};

struct set {
	int capacity;
	int size;
	struct lnode *elems;
};

struct set* new(int capacity) {
	struct set *new_set;
	
	new_set = (struct set*) malloc(sizeof(struct set));
	if(new_set == NULL)
		return NULL; /* no memory left */
	
	new_set->capacity = capacity;
	new_set->size = 0;
	new_set->elems = NULL;
	return new_set;
}

int insert(struct set *s, int x) {
	struct lnode *new_node;
	struct lnode *end_node;
	struct lnode *n;
	
	if(s==NULL)
		return 0; /* NULL set */
		
	if(s->size >= s->capacity)
		return 0; /* no space left */
	
	end_node = s->elems;	
	n = end_node;
	while(n != NULL) {
		if(n->value == x)
			return 0; /* element already in the set */
		end_node = n;
		n = n->next;
	}
	
	/* Creation of new node */
	new_node = (struct lnode*) malloc(sizeof(struct lnode));
	if(new_node == NULL)
		return 0; /* no memory left */
	new_node->value = x;
	new_node->next = s->elems;
	
	s->elems = new_node;
	s->size += 1;
	
	return 1; /* element added */
}

int isnull(struct set *s) {
	if(s==NULL)
		return 1;
	return 0;
}

int isempty(struct set *s) {
	if(s==NULL)
		return 0; 
	if(s->elems==NULL)
		return 1; /* s is empty */
	return 0; 
}

int isfull(struct set *s) {
	if(s==NULL)
		return 0; 
	if(s->size >= s->capacity)
		return 1; /* s is full */
	return 0; 
}

int contains(struct set *s, int x) {
	struct lnode *n;
	
	if(s==NULL)
		return 0; /* s is NULL */
		
	n = s->elems;
	while(n != NULL){
		if(n->value == x)
			return 1; /* element found */
		n = n->next;
	}
	
	return 0; /* element NOT found */
}

int length(struct set *s) {
	struct lnode *n;
	int count;
	
	if(s==NULL)
		return 0; /* s is NULL */
	
	count = 0;
	n = s->elems;
	while(n != NULL){
		count = count + 1;
		n = n->next;
	}
	
	return count;	
}
\end{lstlisting}
\end{multicols}
\end{figure}

%% file: Specs-Base.bbl
\begin{thebibliography}{10}
\providecommand{\url}[1]{\texttt{#1}}
\providecommand{\urlprefix}{URL }

\bibitem{AFV2013}
Alpuente, M., Feli{\'u}, M.A., Villanueva, A.: Automatic {Inference} of
  {Specifications} {Using} {Matching} {Logic}. Proc.\ of PEPM'13, 127--136. ACM (2013)

\bibitem{AFV2015}
Alpuente, M., Pardo, D., Villanueva, A.: Automatic {Inference} of
  {Specifications} in the {K} {Framework}. EPTCS 200: 1--17 (2015)

\bibitem{anand_symbolic_2008}
Anand, S., P{\u a}s{\u a}reanu, C.S., Visser, W.: Symbolic execution with
  abstraction. STTT 11(1):  53--67 (2008)

\bibitem{arusoaie_symbolic_2015}
Arusoaie, A., Lucanu, D., Rusu, V.: Symbolic execution based on language
  transformation. Computer Languages, Systems \& Structures  44, Part A,
  48--71 (2015)

\bibitem{bacci_automatic_2012}
Bacci, G., Comini, M., Feli{\'u}, M.A., Villanueva, A.: 
Automatic {Synthesis}
  of {Specifications} for {First} {Order} {Curry} {Programs}. Proc.\ PPDP'12, 25--34. ACM (2012)

\bibitem{baudin09acsl}
Baudin, P., Filli\^atre, J.C., March\'e, C., Monate, B., Moy, Y., Prevosto, V.:
  ACSL: ANSI/ISO C Specification Language, version 1.4 (2009)

\bibitem{claessen_quickspec_2010} Claessen, K., Smallbone, N., Hughes,
  J.: {QuickSpec}: {Guessing} {Formal} {Specifications} {Using}
  {Testing}. Tests and {Proofs} 2010, Springer LNCS 6143: 6--21 (2010)
  
\bibitem{Cousot2013}
Cousot, P., Cousot, R., F{\"a}hndrich, M., Logozzo, F.: {Automatic Inference of
  Necessary Preconditions}. Proc.\ of VMCAI'13, Springer LNCS 7737: 128--148 (2013)

\bibitem{csallner_dysy_2008}
Csallner, C., Tillmann, N., Smaragdakis, Y.: {DySy}: {Dynamic} {Symbolic}
  {Execution} for {Invariant} {Inference}. Proc.\ of ICSE'08, 281--290.
  ACM (2008)

\bibitem{ellison_executable_2012}
Ellison, C., Ro{\c s}u, G.: An {Executable} {Formal} {Semantics} of {C} with
  {Applications}. Proc.\ of POPL'12, 533--544.
  ACM (2012)

\bibitem{ernst_daikon_2007}
Ernst, M.D., Perkins, J.H., Guo, P.J., McCamant, S., Pacheco, C., Tschantz,
  M.S., Xiao, C.: The {Daikon} system for dynamic detection of likely
  invariants. SCP   69(1{\textendash}3):  35--45
  (2007)

\bibitem{giannakopoulou_interface_2009}
Giannakopoulou, D., P{\u a}s{\u a}reanu, C.S.: Interface {Generation} and
  {Compositional} {Verification} in {JavaPathfinder}. Proc.\ of FASE'09,
  Springer LNCS 5503: 94--108 (2009)


\bibitem{henkel_discovering_2003}
Henkel, J., Diwan, A.: Discovering {Algebraic} {Specifications} from {Java}
  {Classes}. Proc.\ of ECOOP'03. Springer LNCS 2743: 431--456 (2003)

\bibitem{khurshid_generalized_2003}
Khurshid, S., P{\u a}s{\u a}reanu, C.S., Visser, W.: Generalized {Symbolic}
  {Execution} for {Model} {Checking} and {Testing}. Proc.\ of TACAS'03, Springer LNCS 2619: 553--568 (2003)

\bibitem{king_symbolic_1976}
King, J.C.: Symbolic {Execution} and {Program} {Testing}. Comm. ACM  19(7):
  385--394 (1976)

\bibitem{liskov_abstraction_1986}
Liskov, B., Guttag, J.: Abstraction and {Specification} in {Program}
  {Development}. MIT Press, Cambridge, MA, USA (1986)

\bibitem{meyer_applying_1992}
Meyer, B.: Applying 'design by contract'. Computer  25(10):  40--51 (1992)

\bibitem{moura_z3:_2008}
Moura, L.d., Bj{\o}rner, N.: Z3: {An} {Efficient} {SMT} {Solver}. Proc.\ of TACAS'08, Springer LNCS 4963: 337--340 (2008)

\bibitem{rosu_overview_2010}
Ro{\c s}u, G., {\c S}erb{\u a}nu{\c t}{\u a}, T.F.: An overview of the {K}
  semantic framework. JLAP  79(6): 397--434 (2010)

\bibitem{taghdiri_inferring_2006}
Taghdiri, M., Jackson, D.: Inferring specifications to detect errors in code.
  Automated Software Eng.  14(1):  87--121 (2006)

\bibitem{tillmann_discovering_2006}
Tillmann, N., Chen, F., Schulte, W.: Discovering {Likely} {Method}
  {Specifications}. Proc.\ of ICFEM'06, Springer LNCS 4260: 717--736 (2006)

\bibitem{wei_inferring_2011}
Wei, Y., Furia, C.A., Kazmin, N., Meyer, B.: Inferring {Better} {Contracts}.
  Proc.\ of the ICSE'11, 191--200. ACM (2011)

\bibitem{whaley_automatic_2002}
Whaley, J., Martin, M.C., Lam, M.S.: Automatic {Extraction} of
  {Object}-oriented {Component} {Interfaces}. Proc.\ of ISSTA'02, 218--228. ACM (2002)

\end{thebibliography}
